\documentclass[
12pt,
tightenlines,
amsmath,amssymb,
notitlepage
]{revtex4-1}
\pdfoutput=1
\usepackage{graphicx}
\usepackage{dcolumn}
\usepackage{bm}

\usepackage{amsmath}	
\usepackage{ascmac}
\usepackage[colorlinks=true,allcolors=blue]{hyperref}

\begin{document}

\newcommand{\diff}[2]{\frac{d#1}{d#2}}
\newcommand{\pdiff}[2]{\frac{\partial #1}{\partial #2}}
\newcommand{\fdiff}[2]{\frac{\delta #1}{\delta #2}}
\newcommand{\bx}{\bm{x}}
\newcommand{\bX}{\bm{X}}
\newcommand{\bq}{\bm{q}}
\newcommand{\br}{\bm{r}}
\newcommand{\bR}{\bm{R}}
\newcommand{\bu}{\bm{u}}
\newcommand{\by}{\bm{y}}
\newcommand{\bY}{\bm{Y}}
\newcommand{\hv}{\hat{v}}
\newcommand{\hA}{\hat{A}}
\newcommand{\ox}{\overline{x}}
\newcommand{\obx}{\overline{\bm{x}}}
\newcommand{\oy}{\overline{y}}
\newcommand{\oby}{\overline{\bm{y}}}
\newcommand{\new}{\nonumber\\}
\newcommand{\abs}[1]{\left|#1\right|}
\newcommand{\tr}{{\rm Tr}}
\newcommand{\ave}[1]{\left\langle #1 \right\rangle}
\newcommand{\im}{{\rm Im}}
\newcommand{\re}{{\rm Re}}
\newcommand{\erfc}{{\rm erfc}}

\preprint{AIP/123-QED} \title{
Contact-number hyperuniformity at glass close packing in infinite dimensions}

\author{Harukuni Ikeda}
\affiliation{Yukawa Institute for Theoretical Physics, Kyoto University,
Kyoto 606-8502, Japan}

\date{\today}

\begin{abstract}
We investigate contact-number fluctuations of harmonic spheres near
glass close packing (GCP), the amorphous jamming endpoint reached by
compressing an ideal glass. Using replica theory, we show that the
Gardner-unstable region shrinks toward jamming as the spatial dimension
increases, allowing GCP to be approached along a stable one-step
replica symmetry breaking branch in the high-dimensional limit.
By coupling an external field to the local contact number, we derive
the variance of the total contact number and the wave-number-dependent
structure factor directly from the microscopic interactions.
At GCP, the structure factor vanishes quadratically at small scaled
wave numbers, demonstrating contact-number hyperuniformity. Upon weak
compression above GCP, the structure factor develops a low-wave-number
plateau whose height is proportional to the excess coordination.
These results provide an analytical description of
contact-number hyperuniformity and its breakdown upon compression.
\end{abstract}

\maketitle


\section{Introduction}

The jamming transition is characterized by the emergence of mechanical
rigidity upon compression of a particle
assembly~\cite{vanHecke2009,Liu2010,torquato2010}.  Systems of
frictionless repulsive spheres have been extensively studied as a simple
model of this transition~\cite{ohern2003}. Numerical and theoretical
studies have established isostaticity at the jamming transition and
revealed scaling laws for elastic moduli, vibrational properties, and
the distributions of contact forces and interparticle
gaps~\cite{ohern2003,wyart2005,goodrich2012,DeGiuli2014,DeGiuli2014force,Charbonneau2014,Charbonneau2014exact,charbonneau2015,Yan2016,Charbonneau2017,Parisi2019,charbonneau2021,ikeda2022}.

In addition to these scaling laws, numerical studies have revealed an
anomalous suppression of contact-number fluctuations on large length
scales near jamming~\cite{Hexner2018,Hexner2019}. This
behavior is known as contact-number hyperuniformity. In general,
hyperuniformity is characterized by a variance of integrated
fluctuations that grows more slowly than the volume of an observation
window~\cite{Torquato2003,Torquato2018}. For contact-number fluctuations
at jamming, the variance was found to scale with the surface area of the
window rather than its volume~\cite{Hexner2018}. This suppression is
also reflected in the contact-number structure factor $S_z(k)$, which
vanishes with power law $S_z(k)\propto k^\alpha$ as $k\to0$. The
original numerical study reported an approximate scaling $\alpha\approx
1.5$~\cite{Hexner2018,Franz2020,Zhang2026}. More recently, Wang \emph{et al.}\
found $\alpha\simeq 2$ in jammed harmonic-sphere packings with a tuned
particle-size distribution~\cite{Wang2025}.

To explain the origin of contact-number hyperuniformity, Shang \emph{et
al.}\ proposed a minimal network model based on global isostaticity and
local mechanical stability constraints~\cite{Shang2026}. Numerical
simulations of this model yielded contact-number hyperuniformity with an
exponent $\alpha\approx 1.5$. However, to our knowledge, the
contact-number structure factor near jamming has not yet been derived
analytically from a microscopic model of interacting particles.

Replica theory provides a framework for describing glassy states
starting from microscopic particle
interactions~\cite{Monasson1995,parisiZamponi2010}. Early calculations
for harmonic spheres used a one-step replica symmetry breaking (1RSB)
approximation to study glassy and jamming
transitions~\cite{berthier2011}. In the infinite-dimensional limit, the
replicated free-energy functional can be derived
exactly~\cite{Kurchan2012}. As approaches jamming, the 1RSB solution can
become unstable at a Gardner
transition~\cite{Gardner1985,Kurchan2013,Charbonneau2015numerical,Berthier2016,Seguin2016,Biroli2018,Scalliet2019}. Beyond
the Gardner transition, a hierarchy of glassy subbasins emerges,
requiring a full replica symmetry breaking (fullRSB) description. The
fullRSB solution predicts the nontrivial critical scaling of the force
and gap distributions near
jamming~\cite{Charbonneau2014,Charbonneau2014exact,Charbonneau2017,Parisi2019}.
However, calculating spatial contact-number correlations within the
fullRSB framework remains technically challenging.

To make analytical progress, we focus on glass close packing (GCP),
the amorphous jamming endpoint of the ideal-glass
branch~\cite{parisiZamponi2010}. The reduced packing fraction at GCP,
$\hat{\varphi}=2^d\varphi/d$, grows as $\log d$, and the cage size
becomes small~\cite{parisiZamponi2010,Kurchan2013}. The Gardner
boundary moves toward jamming as the reduced density
increases~\cite{Charbonneau2014}. For harmonic spheres, we show that
the width of the Gardner-unstable region, expressed in terms of the
relative excess coordination, decreases as $1/\log d$. In the
limit $d\to\infty$, GCP can thus be approached along a stable 1RSB
branch.

Within this 1RSB description, we introduce an external field
conjugate to the local contact number. The response to a uniform
field yields the mean contact number and the variance of the total
contact number. At GCP, the mean reaches the isostatic value $2d$
and the variance per particle vanishes. Using a spatially varying
field, we derive an explicit expression for the contact-number
structure factor $S_z(q)$. At GCP, it vanishes quadratically in the
scaled wave number $k=\sigma q/\sqrt{d}$, yielding contact-number
hyperuniformity. Upon weak compression above GCP, a finite plateau
develops at small $k$, with a height proportional to the excess
coordination.

The remainder of this paper is organized as follows. In
Sec.~\ref{140415_9Nov26}, we introduce the model and explain how to
compute the mean contact number and its fluctuations.  In
Sec.~\ref{140425_9Nov26}, we derive the replicated free energy in the
high-dimensional and zero-temperature limits.  In
Sec.~\ref{140444_9Nov26}, we analyze the response to a uniform field and
locate the Gardner stability boundary.  In Sec.~\ref{140459_9Nov26}, we
extend the calculation to a spatially varying field and derive the
contact-number structure factor.  We conclude in
Sec.~\ref{140507_9Nov26}.

\section{Setup}
\label{140415_9Nov26}

\subsection{Model}

We consider $N$ harmonic spheres of diameter $\sigma$ in $d$
dimensions~\cite{ohern2003}. The total potential energy is
\begin{align}
H = \sum_{i<j} v(\abs{\bx_i-\bx_j}),
\end{align}
where $\bx_i$ is the position of particle $i$ and the pair potential is
\begin{align}
v(r) = \frac{\kappa}{2}\left(\frac{r}{\sigma}-1\right)^2
\theta(\sigma-r).
\end{align}
Here, $\kappa$ sets the energy scale and $\theta$ is the Heaviside step
function.

\subsection{Contact-number observables}

The contact number of particle $i$ is
\begin{align}
z_i = \sum_{j\neq i}\theta(\sigma-\abs{\bx_i-\bx_j}).
\end{align}
We define the contact field as
\begin{align}
z(\bx) = \sum_{i=1}^N z_i \delta(\bx-\bx_i).
\end{align}
Its Fourier transform is
\begin{align}
z_{\bq} = \int d\bx\, e^{i\bq\cdot\bx}z(\bx)
= \sum_{i=1}^N z_i e^{i\bq\cdot\bx_i}.
\end{align}
The contact-number structure factor is
\begin{align}
S_z(\bq)
= \frac{1}{N}\ave{\delta z_{\bq}\delta z_{-\bq}}
= \frac{\ave{z_{\bq}z_{-\bq}}
- \ave{z_{\bq}}\ave{z_{-\bq}}}{N},
\end{align}
where $\delta z_{\bq}=z_{\bq}-\ave{z_{\bq}}$.
An alternative definition uses the Fourier transformation of
$z_i-\bar z$, where $\bar z$ is the
mean contact number~\cite{Hexner2018}, leading to
\begin{align}
S_z^{'}(\bq) = \frac{1}{N}\ave{\abs{\sum_i e^{i\bq\cdot\bx_i}(z_i-\overline{z})}^2}
 = \frac{1}{N}\left[\ave{z_{\bq}z_{-\bq}}-\overline{z}\ave{z_{\bq}\rho_{-\bq}}
-\overline{z}\ave{z_{-\bq}\rho_{\bq}}
 +\overline{z}^2\ave{\rho_{\bq}\rho_{-\bq}}\right].\label{151727_16Nov26}
\end{align}
The two definitions agree to
leading order in the high-dimensional limit considered here because
density fluctuations contribute only at subleading order
(see Appendix~\ref{154906_16Nov26}).

\subsection{Partition function}

To compute the mean contact number and its correlations, we
introduce a partition function with an external field $\lambda(\bx)$
conjugate to $z(\bx)$,
\begin{align}
Z = \int \prod_{i=1}^N d\bx_i\,
\exp\left[-\beta H + \int d\bx\, z(\bx)\lambda(\bx)\right],
\end{align}
where $\beta$ is the inverse temperature. A functional derivative
with respect to the field gives
\begin{align}
\ave{z(\bx)} = \fdiff{\log Z}{\lambda(\bx)}.
\end{align}
For a uniform field $\lambda(\bx)=\lambda$, the mean contact number
per particle is
\begin{align}
\diff{}{\lambda}\left(\frac{\log Z}{N}\right)
= \frac{1}{N}\ave{\int d\bx\, z(\bx)}
= \frac{1}{N}\sum_{i=1}^N \ave{z_i}.
\end{align}
Using the same Fourier convention for $\lambda(\bx)$ as for $z(\bx)$,
Parseval's identity gives
\begin{align}
\int d\bx\, z(\bx)\lambda(\bx)
= \frac{1}{(2\pi)^d}\int d\bq\, z_{\bq}\lambda_{-\bq}.
\end{align}
Differentiating with respect to the Fourier components of the field
then gives
\begin{align}
\ave{z_{\bq}} = (2\pi)^d \fdiff{\log Z}{\lambda_{-\bq}},\qquad 
\ave{z_{\bq}z_{\bq'}}-\ave{z_{\bq}}\ave{z_{\bq'}}
= (2\pi)^d \fdiff{\ave{z_{\bq}}}{\lambda_{-\bq'}}.
\end{align}

\section{Replica method}
\label{140425_9Nov26}

\subsection{Replicated free energy}

Following Monasson~\cite{Monasson1995}, we consider $m$ replicas
constrained to occupy the same metastable state. The replicated free
energy $F_m$ is defined by
\begin{align}
-\beta F_m = \log Z_m,
\end{align}
where $Z_m$ is the replicated partition function. After analytic
continuation in $m$, the equilibrium liquid corresponds to $m=1$,
whereas the ideal-glass branch has $m<1$. On the glass branch
considered here, $m\to0$ as the temperature goes to zero.

When the replicas occupy the same metastable state, the $m$ copies of
particle $i$ remain localized near one another and can be treated as a
single molecule~\cite{parisiZamponi2010}. We denote its coordinates by
$\obx_i=(\bx_i^1,\ldots,\bx_i^m)$ and write $d\obx_i=\prod_{a=1}^m
d\bx_i^a$ for integration over all replica coordinates of molecule $i$.
The replicated partition function on this glass branch is then written
as
\begin{align}
Z_m = \frac{1}{N!}\int \prod_{i=1}^N d\obx_i\,
\exp\left\{
\sum_{a=1}^m
\left[-\beta H_a+\int d\bx\,z_a(\bx)\lambda_a(\bx)\right]
\right\},
\end{align}
where
\begin{align}
H_a = \sum_{i<j}v(\abs{\bx_i^a-\bx_j^a}), \qquad 
z_a(\bx) = \sum_{i=1}^N z_i^a\delta(\bx-\bx_i^a).
\end{align}
Here, $z_i^a$ is the contact number of particle $i$ in replica $a$.
The source term can be written as a sum over pairs,
\begin{align}
\int d\bx\,z_a(\bx)\lambda_a(\bx)
&= \sum_{i=1}^N z_i^a\lambda_a(\bx_i^a) \notag\\
&= \sum_{i<j}
\left[\lambda_a(\bx_i^a)+\lambda_a(\bx_j^a)\right]
\theta(\sigma-\abs{\bx_i^a-\bx_j^a}).
\end{align}
The Boltzmann weight thus factorizes over molecular pairs, giving
\begin{align}
Z_m = \frac{1}{N!}\int \prod_{i=1}^N d\obx_i\,
\prod_{i<j}\left[1+f(\obx_i,\obx_j)\right],
\end{align}
where the molecular Mayer function is defined by
\begin{align}
f(\obx,\oby)
= \prod_{a=1}^m
e^{
-\beta v(\abs{\bx^a-\by^a})
+\left[\lambda_a(\bx^a)+\lambda_a(\by^a)\right]
\theta(\sigma-\abs{\bx^a-\by^a})}-1.
\end{align}
Following the standard procedure of Morita and
Hiroike~\cite{Morita1960,simpleliquid}, we obtain the replicated
free energy as a functional of the molecular density $\rho(\obx)$,
\begin{align}
-\beta F_m
&= \int d\obx\,\rho(\obx)\left[1-\log\rho(\obx)\right]
\notag\\
&\quad
+\frac{1}{2}\int d\obx\,d\oby\,
\rho(\obx)\rho(\oby)f(\obx,\oby)+\cdots.
\end{align}
The higher-order terms are subleading in the high-dimensional
limit~\cite{parisiZamponi2010,Parisi2019}. The first term is the
ideal-gas contribution $S_{\rm id}$, and the second is the leading
interaction contribution $S_{\rm int}$.

\subsection{Gaussian ansatz}

Tomita and Yoshino developed a replicated liquid theory in which the
density and the glass order parameter vary along a single spatial
direction~\cite{Tomita2025}. Here we allow the density and cage
variances to depend on the full molecular-center coordinate $\bX$. We
use a Gaussian ansatz for the molecular density,
\begin{align}
\rho(\obx) = \int d\bX\,n(\bX)
\prod_{a=1}^m \gamma_{A_a(\bX)}(\bx^a-\bX),
\end{align}
where $\gamma_A$ is a normalized Gaussian with variance $A$ per
Cartesian component. The local density of molecular centers,
$n(\bX)$, is normalized to the total number of molecules,
\begin{align}
\int d\bX\,n(\bX)=N.
\end{align}
Since the source $\lambda_a(\bx)$ can differ between replicas, we
initially allow the cage variance $A_a(\bX)$ to depend on the replica
index $a$. In the zero-temperature limit, however, the saddle-point
equations admit a solution with the same rescaled cage variance for all
replicas at each $\bX$, as we will see below.

\subsection{High-dimensional limit}

Substituting the Gaussian ansatz for the density outside the
logarithm gives
\begin{align}
S_{\rm id} &\equiv \int d\obx\, \rho(\obx)\left[1-\log\rho(\obx)\right]\new 
&= \int d\bX\,n(\bX)\int d\obx\,
\prod_{a=1}^m \gamma_{A_a(\bX)}(\bx^a-\bX)
\left[1-\log\rho(\obx)\right].
\end{align}
The logarithm contains a separate integral over the molecular center,
\begin{align}
\log\rho(\obx)
= \log\left[
\int d\bY\,n(\bY)
\prod_{a=1}^m \gamma_{A_a(\bY)}(\bx^a-\bY)
\right].
\end{align}
At large $d$, the cage variances are of order
$d^{-2}$~\cite{parisiZamponi2010}. At fixed $\bX$, the outer Gaussian
factors therefore localize the replica coordinates $\bx^a$ around the
molecular center $\bX$. The Gaussian product inside the logarithm is
appreciable when $\bY$ is close to these replica coordinates, so the
dominant contribution to the inner integral comes from $\bY$ near
$\bX$. We thus approximate $n(\bY)$ and $A_a(\bY)$ by $n(\bX)$ and
$A_a(\bX)$, respectively, in the inner integral,
\begin{align}
\log\rho(\obx)
\simeq \log\left[
n(\bX)\int d\bY\,
\prod_{a=1}^m \gamma_{A_a(\bX)}(\bx^a-\bY)
\right].
\end{align}
We perform the Gaussian integral over $\bY$ and average the resulting
logarithm over the replica coordinates at fixed $\bX$. To express
the result, we introduce the
dimensionless variances $\hA_a(\bX)$ through
\begin{align}
A_a(\bX)=\frac{\sigma^2}{d^2}\hA_a(\bX).
\end{align}
The ideal contribution then reads~\cite{parisiZamponi2010}
\begin{align}
S_{\rm id}
&\simeq \int d\bX\,n(\bX)
\left\{
1-\log n(\bX)
+\frac{d}{2}\left[
(m-1)\log\left(\frac{2\pi e\,\sigma^2}{d^2}\right)
+\sum_{a=1}^m \log \hA_a(\bX)
+\log\left(\sum_{a=1}^m\frac{1}{\hA_a(\bX)}\right)
\right]
\right\}.
\end{align}

Substituting the Gaussian ansatz into the interaction contribution
gives
\begin{align}
S_{\rm int} &\equiv \frac{1}{2}\int d\obx\, d\oby\, \rho(\obx)\rho(\oby)f(\obx,\oby)\new 
&= \frac{1}{2}\int d\bX\,d\bY\,n(\bX)n(\bY)
\left[\prod_{a=1}^m Q_a(\bX,\bY)-1\right],
\end{align}
where
\begin{align}
Q_a(\bX,\bY)
&= \int d\bx\,d\by\,
\gamma_{A_a(\bX)}(\bx-\bX)
\gamma_{A_a(\bY)}(\by-\bY) \notag\\
&\quad\times
\exp\left[
-\beta v(\abs{\bx-\by})
+\left[\lambda_a(\bx)+\lambda_a(\by)\right]
\theta(\sigma-\abs{\bx-\by})
\right].
\end{align}
When the sources vary slowly on the cage scale, we evaluate
$\lambda_a(\bx)$ and $\lambda_a(\by)$ at the corresponding molecular
centers $\bX$ and $\bY$. We define the scaled gap and inverse
temperature as
\begin{align}
h=d\left(\frac{\abs{\bX-\bY}}{\sigma}-1\right),
\qquad
\hat{\beta}=\frac{\beta \kappa}{d^2}.
\end{align}
At fixed $h$ in the high-dimensional limit, we
obtain~\cite{parisiZamponi2010,Biroli2018}
\begin{align}
Q_a(\bX,\bY)
&\simeq \int ds\,\gamma_{\hA_a(\bX)+\hA_a(\bY)}(s)
\exp\left\{
\left[-\frac{\hat{\beta}}{2}g_a^2
+\lambda_a(\bX)+\lambda_a(\bY)\right]\theta(-g_a)
\right\}, \notag\\
g_a &= h-s+\frac{\hA_a(\bX)+\hA_a(\bY)}{2}.
\end{align}

\subsection{Zero-temperature limit}

We take the joint limit $\hat{\beta}\to\infty$ and $m\to0$ at fixed
ratio, writing
\begin{align}
\hA_a(\bX)=m\alpha_a(\bX),\qquad
\hat{\beta}=\frac{1}{m\tau}.
\end{align}
For fixed $h$, the saddle point method leads to 
\begin{align}
\prod_{a=1}^m Q_a(\bX,\bY)
&\simeq \theta(h)+\theta(-h)
\exp\left[
\Lambda(\bX)+\Lambda(\bY)
-\frac{h^2}{2m}\sum_{a=1}^m
\frac{1}{\tau+\alpha_a(\bX)+\alpha_a(\bY)}
\right], 
\end{align}
where
\begin{align}
\Lambda(\bX) &\equiv \sum_{a=1}^m\lambda_a(\bX).
\end{align}
The sources enter the free-energy only
through their sum $\Lambda(\bX)=\sum_{a=1}^m\lambda_a(\bX)$. The
free-energy is thus invariant under permutations of the replica
labels on the cage fields $\alpha_a(\bX)$. We seek a saddle that
preserves this symmetry,
\begin{align}
\alpha_a(\bX)=\alpha(\bX),\qquad a=1,\ldots,m.
\end{align}
With this choice, the stationarity conditions for the individual
replicas reduce to a single equation for $\alpha(\bX)$. At this saddle,
taking $m\to0$ in the ideal term gives
\begin{align}
S_{\rm id}
= \int d\bX\,n(\bX)
\left[
1-\log n(\bX)
-\frac{d}{2}\log\left(\frac{2\pi e\,\sigma^2\alpha(\bX)}{d^2}\right)
\right],
\label{152446_4Nov26}
\end{align}
and the interaction term becomes
\begin{align}
S_{\rm int}
&= \frac{1}{2}\int d\bX\,d\bY\,n(\bX)n(\bY)\theta(-h)
\left\{
\exp\left[
\Lambda(\bX)+\Lambda(\bY)
-\frac{h^2}{2[\tau+\alpha(\bX)+\alpha(\bY)]}
\right]-1
\right\}.
\label{152454_4Nov26}
\end{align}

\section{Homogeneous case}
\label{140444_9Nov26}

\subsection{Free energy}

We first consider a spatially uniform source,
$\lambda_a(\bx)=\lambda_a$. The homogeneous saddle has
$n(\bX)=n$ and $\alpha(\bX)=\alpha$. From
Eq.~\eqref{152446_4Nov26}, the ideal contribution per particle is
\begin{align}
\frac{S_{\rm id}}{N}
= 1-\log n
+\frac{d}{2}\log\left(\frac{d^2}{2\pi e\,\sigma^2}\right)
-\frac{d}{2}\log\alpha.
\end{align}
Let $v_d=\pi^{d/2}\sigma^d/\Gamma(d/2+1)$ be the volume of a
$d$-dimensional ball of radius $\sigma$.  Then
$nv_d=2^d\varphi=d\hat{\varphi}$, where $\hat{\varphi}=2^d\varphi/d$.
At the GCP, $\hat{\varphi}=O(\log d)$~\cite{parisiZamponi2010}.  Using
Stirling's formula for $v_d$, we obtain
\begin{align}
\frac{S_{\rm id}}{N}
= -\frac{d}{2}\log\alpha+\frac{d}{2}\log d
+O(\log d).
\end{align}

After angular integration, the radial measure in
Eq.~\eqref{152454_4Nov26} becomes $d\hat{\varphi}\,e^h\,dh$ in the
high-dimensional limit. With $\Lambda=\sum_a\lambda_a$, the
interaction contribution per particle is
\begin{align}
\frac{S_{\rm int}}{N}
= \frac{d\hat{\varphi}}{2}
\left[e^{2\Lambda}G\left(\alpha+\frac{\tau}{2}\right)-1\right],
\end{align}
where
\begin{align}
G(x)
= \int_{-\infty}^{0}dh\,e^{h-h^2/(4x)}
= \sqrt{\pi x}\,e^x\operatorname{erfc}(\sqrt{x}).
\end{align}
For small $x$, this function has the expansion
\begin{align}
G(x)=\sqrt{\pi x}-2x+O(x^{3/2}).
\end{align}

\subsection{Cage size} 
The cage size is calculated by minimizing the free-energy
\begin{align}
s \equiv \frac{S_{\rm id}+S_{\rm int}}{N}.
\end{align}
The cage size is determined by the saddle-point condition:
\begin{align}
0 = \frac{\partial s}{\partial\alpha}
= -\frac{d}{2\alpha}
+\frac{d\hat{\varphi}}{2}e^{2\Lambda}
G'\left(\alpha+\frac{\tau}{2}\right),
\end{align}
which gives
\begin{align}
\frac{1}{\hat{\varphi}}
= e^{2\Lambda}\alpha G'\left(\alpha+\frac{\tau}{2}\right).
\label{120005_3Nov26}
\end{align}
We introduce the dimensionless ratio
\begin{align}
\varepsilon\equiv\frac{\tau}{2\alpha}.
\end{align}
For weak compression along the ideal-glass branch in the
high-dimensional limit, $\varepsilon$ increases linearly with
compression above GCP and measures the relative excess coordination, as
we will show in the following subsections. Using
$G'(x)\simeq\sqrt{\pi}/(2\sqrt{x})$ in Eq.~\eqref{120005_3Nov26}, we
find
\begin{align}
\alpha \simeq
\frac{4(1+\varepsilon)e^{-4\Lambda}}{\pi\hat{\varphi}^{\,2}}.
\end{align}
Thus, at fixed $\varepsilon$ and $\Lambda$, the cage size decreases as
$\alpha\sim \hat{\varphi}^{-2}\sim (\log d)^{-2}$, which validates the small-cage expansion at large $d$.

\subsection{Complexity}

The zero-complexity condition selects the ideal-glass branch.  Together
with the saddle-point equation, it determines $\tau$ and hence
$\varepsilon$ as functions of $\hat{\varphi}$, as we will see below.  At
zero source, the complexity per particle is
\begin{align}
\Sigma=s_{m}
-m\frac{\partial s_{m}}{\partial m},
\end{align}
where $s_{m}$ is the replicated entropy per particle at finite
$m$~\cite{Monasson1995,parisiZamponi2010}. Writing $\hat T=m\tau$ and
taking the zero-temperature limit gives~\cite{berthier2011}
\begin{align}
\Sigma(\tau)=s(\tau)
+\tau\frac{\partial s}{\partial\tau}.
\end{align}
To reach the jamming endpoint, we first take $\hat T\to0$ at fixed
$m>0$. This suppresses particle overlaps and reduces the model to hard
spheres. We then take $m\to0$ along the hard-sphere glass branch,
reaching its infinite-pressure jamming
endpoint~\cite{parisiZamponi2010}. This order of limits corresponds to
$\tau=\hat T/m=0$, for which $\Sigma(0)=s(0)$.  In the density regime
$\hat{\varphi}=O(\log d)$, the leading terms are
\begin{align}
\Sigma(0)
= \frac{d}{2}\log d-\frac{d\hat{\varphi}}{2}
+O(d\log\log d).
\label{224852_2Nov26}
\end{align}
Setting $\Sigma(0)=0$ gives the GCP density,
\begin{align}
\hat{\varphi}_J^{\rm GCP}=\log d+O(\log\log d),
\end{align}
consistent with Ref.~\cite{parisiZamponi2010}. 

To determine how $\varepsilon$ changes upon compression above GCP, we
expand the complexity at small $\tau$. The stationarity of $s$ with
respect to $\alpha$ gives
\begin{align}
\left.\frac{\partial s}{\partial\tau}\right|_{\hat{\varphi}}
= \frac{d\hat{\varphi}}{4}G'\left(\alpha+\frac{\tau}{2}\right)
= \frac{d}{4\alpha},
\end{align}
where the last equality follows from Eq.~\eqref{120005_3Nov26} at
zero source. With $\alpha_0=\alpha(\tau=0)$, we obtain
\begin{align}
\Sigma(\tau)-\Sigma(0)
&= \frac{\tau d}{2\alpha_0}
+O\!\left(d\left(\frac{\tau}{\alpha_0}\right)^2\right)
\notag\\
&= d\varepsilon+O(d\varepsilon^2).
\end{align}
At $\tau=0$, differentiation with respect to density gives
\begin{align}
\frac{d\Sigma(0)}{d\hat{\varphi}}
= -\frac{1}{\hat{\varphi}}
+\frac{d}{2}\left[G(\alpha_0)-1\right]
= -\frac{d}{2}+O\!\left(\frac{d}{\hat{\varphi}}\right).
\end{align}
Expanding about $\hat{\varphi}_J^{\rm GCP}$ and imposing
$\Sigma(\tau)=0$, we find
\begin{align}
\varepsilon \simeq
\frac{\hat{\varphi}-\hat{\varphi}_J^{\rm GCP}}{2}
\end{align}
for weak compression above GCP.

\subsection{Gardner transition}

At zero source, we locate the Gardner transition of the 1RSB
solution using the condition derived by Biroli and
Urbani~\cite{Biroli2018},
\begin{align}
1
&= \frac{\hat{\varphi}}{2}e^{-\hA}
\int_{-\infty}^{\infty}dh\,e^h Q(h)^m
\left[2\hA\frac{d^2}{dh^2}\log Q(h)\right]^2,
\label{115816_3Nov26}
\end{align}
where
\begin{align}
Q(h)
= \int dz\,\gamma_{2\hA}(z)
\exp\left[-\frac{\hat{\beta}}{2}(h-z)^2\theta(z-h)\right].
\end{align}
We set $\hA=m\alpha$ and $\hat{\beta}=1/(m\tau)$. In the limit
$m\to0$ at fixed $\tau>0$, we obtain, for $h\ne0$,
\begin{align}
Q(h)^m
\longrightarrow
\exp\left[-\frac{h^2}{4(\alpha+\tau/2)}\right]\theta(-h)
+\theta(h), \qquad 
2\hA\frac{d^2}{dh^2}\log Q(h)
\longrightarrow -\frac{\alpha}{\alpha+\tau/2}\theta(-h).
\end{align}
Substituting these limits into Eq.~\eqref{115816_3Nov26} gives
\begin{align}
1
= \frac{\hat{\varphi}}{2}
\frac{\alpha^2}{(\alpha+\tau/2)^2}
G\left(\alpha+\frac{\tau}{2}\right).
\end{align}
Using the zero-source saddle-point condition
\eqref{120005_3Nov26}, we eliminate $\hat{\varphi}$ to obtain
\begin{align}
1
= \frac{\alpha}{2(\alpha+\tau/2)^2}
\frac{G(\alpha+\tau/2)}{G'(\alpha+\tau/2)}.
\label{135650_4Nov26}
\end{align}
For small $x$, the ratio in this equation has the expansion
\begin{align}
\frac{G(x)}{G'(x)}
=2x\left[1+2\sqrt{\frac{x}{\pi}}+O(x)\right].
\end{align}
Using $\alpha+\tau/2=\alpha(1+\varepsilon)$, we find that the
Gardner boundary occurs at
\begin{align}
\varepsilon_{\rm G}
= 2\sqrt{\frac{\alpha_{\rm G}}{\pi}}+O(\alpha_{\rm G}),
\end{align}
where $\alpha_{\rm G}$ is the cage size at the transition. Near
GCP, $\alpha_{\rm G}=O((\log d)^{-2})$, so the width of the unstable
region is
\begin{align}
\varepsilon_{\rm G}=O\!\left(\frac{1}{\log d}\right).
\end{align}
The Gardner boundary thus approaches jamming as $d$ increases.  At fixed
$\varepsilon>0$ near GCP, the 1RSB solution is stable against the
Gardner instability in the limit $d\to\infty$.  The jamming endpoint
$\varepsilon=0$ is marginal in this limit~\cite{Kurchan2013}.

\begin{figure}[t]
\centering \includegraphics[width=\linewidth]{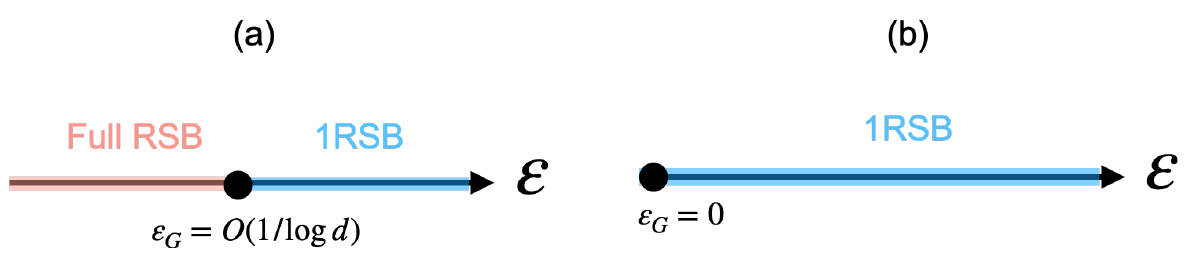}
\caption{Schematic stability diagram of the 1RSB solution for harmonic
spheres near GCP. The parameter $\varepsilon=\tau/(2\alpha)$ measures
the relative excess coordination $(\bar z-2d)/(2d)$ to leading order at
large $d$.  (a) At large but finite $d$, the 1RSB solution is unstable
for $0<\varepsilon<\varepsilon_{\rm G}$, with $\varepsilon_{\rm
G}=O(1/\log d)$.  (b) As $d\to\infty$, the Gardner boundary approaches
the marginal jamming endpoint $\varepsilon=0$, and the 1RSB solution is
stable at fixed $\varepsilon>0$ near GCP.}
\end{figure}

\subsection{Mean contact number and its fluctuations}

The uniform source $\lambda_a$ couples to the sum of the particle
contact numbers in replica $a$:
\begin{align}
\int d\bx z_a(\bx)\lambda_a(\bx) = \lambda_a Z_a, 
\end{align}
where 
\begin{align}
Z^a=\sum_{i=1}^N z_i^a.
\end{align}
We obtain the mean contact number per particle by
differentiating the free-energy $s$ with respect to $\lambda_a$.
Since $s$ is stationary with respect
to $\alpha$, only the explicit source dependence contributes to
the first derivative,
\begin{align}
\bar z_a
= \frac{1}{N}\ave{Z^a}
= \frac{\partial s}{\partial\lambda_a}
= d\hat{\varphi}e^{2\Lambda}G\left(\alpha+\frac{\tau}{2}\right).
\end{align}
Using Eq.~\eqref{120005_3Nov26}, we can also write
\begin{align}
\bar z_a
= \frac{d}{\alpha}
\frac{G(\alpha+\tau/2)}{G'(\alpha+\tau/2)}.
\end{align}
The leading small-cage result is
\begin{align}
\bar z \simeq 2d\left(1+\frac{\tau}{2\alpha}\right)
=2d(1+\varepsilon).
\label{174720_2Nov26}
\end{align}
Thus, near GCP in the high-dimensional limit, $\varepsilon$ measures the
relative excess coordination
\begin{align}
\varepsilon\simeq \frac{\bar{z}-2d}{2d}.
\end{align}
At $\varepsilon=0$, the limiting mean contact number reaches the
isostatic value $2d$. Combining Eq.~\eqref{174720_2Nov26} with the
zero-complexity condition gives a linear increase upon weak compression
in the high-dimensional 1RSB regime,
\begin{align}
\bar z-2d\simeq d(\hat{\varphi}-\hat{\varphi}_J^{\rm GCP}).
\end{align}
In finite-dimensional simulations of harmonic spheres, the excess
coordination grows as $\bar z-2d\propto(\varphi-\varphi_J)^{1/2}$ near
jamming~\cite{ohern2003}.  The difference between these scaling laws
reflects the replica structure governing the approach to jamming. At
GCP, where $\hat{\varphi}_J^{\rm GCP}\sim\log d$, the Gardner-unstable
region shrinks as $d\to\infty$, allowing jamming to be approached along
a stable 1RSB branch. At finite $d$, a glass can undergo a Gardner
transition before reaching jamming. This can also occur in the limit
$d\to\infty$ when the jamming density $\hat{\varphi}_J$ remains
finite~\cite{Biroli2018,Scalliet2019}. In either case, full replica
symmetry breaking governs the asymptotic scaling near
jamming~\cite{Charbonneau2014exact,Charbonneau2014}.

\begin{figure}[t]
\centering
\includegraphics[width=10cm]{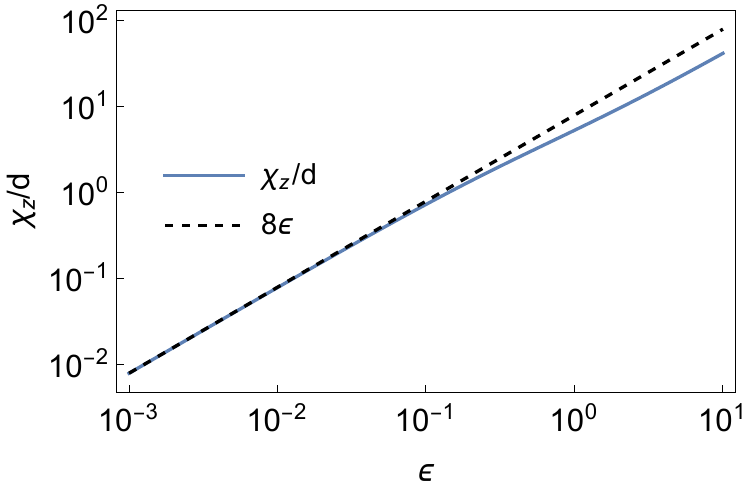}
\caption{Contact-number susceptibility
$\chi_z=N^{-1}\operatorname{Var}(Z^a)$ as a function of
$\varepsilon=\tau/(2\alpha)$.
The parameter $\varepsilon=\tau/(2\alpha)$ measures
the relative excess coordination $(\bar z-2d)/(2d)$ to leading order at
large $d$. The solid line shows the leading
1RSB result and the dashed line shows its small-$\varepsilon$
asymptote, $\chi_z\simeq8d\varepsilon$.}
\label{222130_2Nov26}
\end{figure}

Further derivative with respect to $\lambda_a$ gives the
contact-number susceptibility,
\begin{align}
\chi_z
= \frac{1}{N}\operatorname{Var}(Z^a)
= \frac{\partial\bar z_a}{\partial\lambda_a}.
\end{align}
Using the leading small-cage expression
$\bar z_a\simeq 2d(1+\varepsilon)$ and differentiating at fixed
$\hat{\varphi}$ and $\tau$, we obtain
\begin{align}
\chi_z
\simeq 2d\frac{\partial\varepsilon}{\partial\lambda_a}
= -2d\varepsilon
\frac{\partial\log\alpha}{\partial\lambda_a}.
\label{175850_2Nov26}
\end{align}
Taking the logarithm of the small-cage form of
Eq.~\eqref{120005_3Nov26} and differentiating with respect to
$\lambda_a$ gives
\begin{align}
0 \simeq
2+\left[1-\frac{1}{2(1+\varepsilon)}\right]
\frac{\partial\log\alpha}{\partial\lambda_a}.
\end{align}
Substituting this result into Eq.~\eqref{175850_2Nov26} yields
\begin{align}
\chi_z \simeq \frac{8d\varepsilon(1+\varepsilon)}{1+2\varepsilon}.
\end{align}
Figure~\ref{222130_2Nov26} shows this result. As jamming is approached,
$\chi_z\simeq 8d\varepsilon\simeq 4\delta z$, where $\delta z\equiv
\bar{z}-2d$ denotes the excess coordination. The 1RSB calculation thus
gives
\begin{align}
\chi_z \propto (\delta z)^\beta,\qquad \beta=1.
\end{align}
Numerical simulations in three dimensions suggest $\beta\simeq
1.2$--$1.25$~\cite{Hexner2019,Giannini2024,Shiraishi2026}.
Whether extending our calculation to the fullRSB regime yields an
exponent closer to these estimates remains an open question.

\section{Inhomogeneous case}
\label{140459_9Nov26}

In this section, we extend the calculation to
a spatially varying source to determine how the contact-number
fluctuations depend on wave number.

\subsection{Free energy}

We apply a spatially varying source with zero spatial average and
expand Eqs.~\eqref{152446_4Nov26} and \eqref{152454_4Nov26} about
the homogeneous saddle at zero source. We write the density and
cage fields as
\begin{align}
n(\bX)=n[1+r(\bX)],
\qquad
\alpha(\bX)=\alpha[1+a(\bX)].
\end{align}
For a source at nonzero wave number, the linear responses $r$ and
$a$ have zero spatial average.

Near GCP, $\alpha=O((\log d)^{-2})$ and, at fixed
$\varepsilon=\tau/(2\alpha)$, $\tau$ is of the same order. The Gaussian
in the interaction term is therefore confined to a narrow shell on the
inside of the contact surface $\abs{\bX-\bY}=\sigma$. To leading order
in the shell width~\cite{Mangeat2016},
\begin{align}
&\exp\left[-\frac{h^2}{2[\tau+\alpha(\bX)+\alpha(\bY)]}\right]
\theta(-h)
\simeq
\sqrt{\frac{\pi}{2}\bigl[\tau+\alpha(\bX)+\alpha(\bY)\bigr]}\,
\frac{\sigma}{d}\,
\delta\!\left(\abs{\bX-\bY}-\sigma\right).
\end{align}
The square-root factor is the integrated weight of the Gaussian
over $h<0$, and $\sigma/d$ converts the delta function in $h$ to
one in the radial distance.

We use $\ave{f}_V=V^{-1}\int d\bX\,f(\bX)$ for spatial averages.
For functions of two positions, we define averages over the surface
and interior of the contact sphere,
\begin{align}
\ave{f}_{s_d}
&=\frac{1}{V s_d}\int d\bX\,d\bY\,
\delta\!\left(\abs{\bX-\bY}-\sigma\right)f(\bX,\bY), \notag\\
\ave{f}_{v_d}
&=\frac{1}{V v_d}\int d\bX\,d\bY\,
\theta\!\left(\sigma-\abs{\bX-\bY}\right)f(\bX,\bY),
\end{align}
where $s_d=dv_d/\sigma$ is the surface area of a sphere of radius
$\sigma$. With ${r_X=r(\bX)}$, ${a_X=a(\bX)}$, and
${\Lambda_X=\Lambda(\bX)}$, the ideal and
interaction contributions are, up to additive constants,
\begin{align}
&\frac{S_{\rm id}}{N}
=-\ave{(1+r_X)\log(1+r_X)}_V
-\frac{d}{2}\ave{(1+r_X)\log(1+a_X)}_V,\\
&\frac{S_{\rm int}}{N}
\simeq \frac{\bar z}{2}
\ave{(1+r_X)(1+r_Y)e^{\Lambda_X+\Lambda_Y}
\sqrt{1+\frac{a_X+a_Y}{2(1+\varepsilon)}}}_{s_d}-\frac{d\hat{\varphi}}{2}\ave{r_Xr_Y}_{v_d}.
\end{align}
Here $\bar z\simeq d\hat{\varphi}\sqrt{\pi\alpha(1+\varepsilon)}
\simeq 2d(1+\varepsilon)$ is the mean contact number of the
homogeneous saddle to leading order.

For a modulation at fixed nonzero scaled wave number
$k=\sigma|\bq|/\sqrt{d}$, the excluded-volume term
$-d\hat{\varphi}\ave{r_Xr_Y}_{v_d}/2$ gives a density stiffness of
order $d\hat{\varphi}=O(d\log d)$ near GCP. The couplings of the
density mode to the cage modulation and the contact-number source
are both of order $d$. The density saddle point therefore gives
\begin{align}
r_X
=O\!\left(\frac{a_X}{\log d}, \frac{\Lambda_X}{\log d}\right)
\end{align}
so that we can set $r_X=0$ when computing the leading-order
contact-number response.

Setting $r_X=0$ and expanding to second order in the cage modulation
$a_X$ and the source $\Lambda_X$, we obtain
\begin{align}
\frac{S_{\rm id}^{(2)}}{N}
&=\frac{d}{4}\ave{a_X^2}_V, \notag\\
\frac{S_{\rm int}^{(2)}}{N}
&=\frac{d(1+\varepsilon)}{2}
\ave{(\Lambda_X+\Lambda_Y)^2}_{s_d}
-\frac{d}{32(1+\varepsilon)}
\ave{(a_X+a_Y)^2}_{s_d} \notag\\
&\quad+\frac{d}{4}
\ave{(a_X+a_Y)(\Lambda_X+\Lambda_Y)}_{s_d}.
\end{align}
With the Fourier convention introduced above, the surface average
of two fields $f$ and $g$ takes the form
\begin{align}
\ave{(f_X+f_Y)(g_X+g_Y)}_{s_d}
=\frac{2}{(2\pi)^d V}\int d\bq\,
(1+K_q)f_{\bq}g_{-\bq},
\end{align}
where
\begin{align}
K_q=\frac{1}{s_d}\int d\bx\,e^{i\bq\cdot\bx}
\delta\!\left(\abs{\bx}-\sigma\right)
\end{align}
is the Fourier transform of the normalized surface measure on the
contact sphere. In the limit $d\to\infty$, it converges to
\begin{align}
K_q \xrightarrow{d\to\infty} \exp\left[-\frac{k^2}{2}\right],\qquad k=\frac{\sigma q}{\sqrt{d}}.\label{173037_13Nov26}
\end{align}
The quadratic contribution to $-\beta F_m$ is then
\begin{align}
(-\beta F_m)^{(2)}
=\frac{N}{(2\pi)^d V}\int d\bq\,
\bigl[
A_q a_{\bq}a_{-\bq}
+B_q a_{\bq}\Lambda_{-\bq}
+C_q\Lambda_{\bq}\Lambda_{-\bq}
\bigr],
\end{align}
with
\begin{align}
A_q &= \frac{d}{4}-\frac{d(1+K_q)}{16(1+\varepsilon)}, \notag\\
B_q &= \frac{d}{2}(1+K_q), \notag\\
C_q &= d(1+\varepsilon)(1+K_q).
\end{align}

\subsection{Contact-number structure factor}

\begin{figure}[t]
\centering \includegraphics[width=10cm]{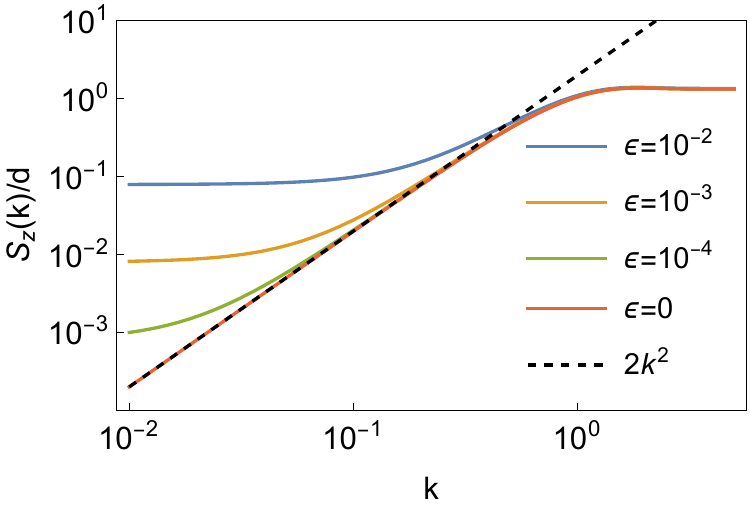} \caption{Contact-number
structure factor $S_z(q)/d$ as a function of the scaled wave number
$k=\sigma q/\sqrt{d}$ for several values of
$\varepsilon=\tau/(2\alpha)$.  The parameter
$\varepsilon=\tau/(2\alpha)$ measures the relative excess coordination
$(\bar z-2d)/(2d)$ to leading order at large $d$. The small-$k$ plateau
decreases as $\varepsilon$ approaches zero. At GCP ($\varepsilon=0$),
the structure factor vanishes quadratically at small $k$.}
\label{183827_20Oct26}
\end{figure}

Stationarity with respect to $a_{-\bq}$ gives the cage response,
\begin{align}
a_{\bq}=-\frac{B_q}{2A_q}\Lambda_{\bq}.
\end{align}
Substituting this response into the quadratic free energy gives
\begin{align}
(-\beta F_{m,*})^{(2)}
= \frac{n}{(2\pi)^d}\int d\bq\,
\left(C_q-\frac{B_q^2}{4A_q}\right)
\Lambda_{\bq}\Lambda_{-\bq},
\end{align}
where $F_{m,*}$ denotes the value at the saddle. Since each replica
source enters through $\Lambda$, differentiating twice with respect
to the source in one replica yields the connected correlation
function
\begin{align}
\ave{\delta z_{\bq}\delta z_{\bq'}}
\simeq (2\pi)^d n
\left(2C_q-\frac{B_q^2}{2A_q}\right)
\delta(\bq+\bq').
\end{align}
In a periodic volume $V$, the Dirac delta is replaced by
$V/(2\pi)^d$ times the corresponding Kronecker delta. Dividing
the covariance at $\bq'=-\bq$ by $N$ therefore gives
\begin{align}
S_z(q)
&\simeq 2C_q-\frac{B_q^2}{2A_q} \notag\\
&=4d(1+\varepsilon)
\frac{(1+K_q)(1+2\varepsilon-K_q)}
{3+4\varepsilon-K_q}.
\end{align}
For fixed $k=\sigma q/\sqrt{d}$ in the high-dimensional limit, 
using Eq.~(\ref{173037_13Nov26}), we get
\begin{align}
S_z(k) = 4d(1+\varepsilon)
\frac{(1+e^{-k^2/2})(1+2\varepsilon-e^{-k^2/2})}
{3+4\varepsilon-e^{-k^2/2}}.
\end{align}
Figure~\ref{183827_20Oct26} shows the resulting structure factor
as a function of $k$. The plateau at small $k$ decreases as
jamming is approached. Taking $k\to0$ in the limiting spectrum
gives
\begin{align}
\lim_{k\to0}S_z(k)
=\chi_z.
\end{align}
At GCP, where $\chi_z=0$, the plateau vanishes
and the leading small-$k$ behavior is
\begin{align}
S_z(k)/d = 2k^2 + O(k^4).
\end{align}
The quadratic decay establishes contact-number hyperuniformity at GCP in
the high-dimensional limit. The exponent $2$ is larger than the value of
about $1.5$ reported by Hexner \emph{et al.}~\cite{Hexner2018}. Whether
full replica symmetry breaking changes the wave-number exponent in the
marginal regime remains an open question.

\section{Summary and discussion}
\label{140507_9Nov26}

We derived the replicated free-energy functional of harmonic spheres
near glass close packing (GCP) in the high-dimensional limit within the
1RSB ansatz. By coupling a source to the contact field, we obtained both
the uniform susceptibility and the wave-number-dependent structure
factor. For a uniform source, the mean contact number $\bar z$ reaches
the isostatic value $2d$ at GCP and increases linearly with the excess
packing fraction $\hat{\varphi}-\hat{\varphi}_J^{\rm GCP}$ upon compression. The variance of the total
contact number per particle, $\chi_z$, vanishes at the transition point
as $\chi_z \propto \delta z^\beta$ with $\beta=1$.
For a spatially varying source, we obtained the contact-number
structure factor. Taking the high-dimensional limit at
fixed scaled wave number $k=\sigma q/\sqrt{d}$ gives
\begin{align}
\frac{S_z(k)}{d}\simeq \frac{4\delta z}{d} +2k^2,\quad \frac{\delta z}{d}\ll 1,\ k\ll 1.
\end{align}
At GCP, $\delta z=0$ and the contact number exhibits hyperuniformity
$S_z(k)\propto k^\alpha$ with $\alpha=2$. Above GCP, $\delta z>0$, and
thus, hyperuniformity is not observed.

Our results $\beta=1$ and $\alpha=2$ deviate from those of
three-dimensional simulations, which reported
$\alpha\simeq1.5$~\cite{Hexner2018} and
$\beta\simeq1.2$--$1.25$~\cite{Hexner2019,Giannini2024,Shiraishi2026}.
The difference may be related to the stability of the glass on the
approach to jamming. At high-dimensional GCP, the width of the
Gardner-unstable region scales as $1/\log d$. In the limit $d\to\infty$,
this region shrinks to the jamming point, which can then be approached
along a stable 1RSB branch and is marginal at its endpoint. A direction
for future work is to extend the current calculation to the fullRSB
regime and determine whether the hierarchy of glass subbasins yields
values of $\alpha$ and $\beta$ closer to the numerical results in finite
dimension.

Wang \emph{et al.}\ found $\alpha\simeq 2$ in jammed harmonic-sphere
packings with a tuned particle-size distribution~\cite{Wang2025}, in
agreement with our 1RSB prediction. They also found enhanced vibrational
stability in glasses prepared by particle resizing well above jamming,
where low-frequency quasilocalized modes were strongly suppressed. This
enhanced stability raises the possibility that these glasses approach
jamming along a stable 1RSB branch, which could explain the agreement in
$\alpha$. Further work is needed to elucidate this point.

\begin{acknowledgments}
This project has received JSPS KAKENHI Grant Numbers 23K13031, and
25H01401.
 
\par\noindent\textbf{AI disclosure.}
I used ChatGPT--6.1 (OpenAI) to assist with drafting and editing the
manuscript, and checking intermediate steps in the derivations. I
checked the derivations against my calculations, and reviewed and
revised the AI-assisted text. I take full responsibility for the content
of the manuscript.
\end{acknowledgments}

\section*{Data availability}

This study is based on analytical calculations. All quantitative
plots can be reproduced from the expressions given in the article.

\appendix 
\section{Comparison of the two contact-number structure factors}
\label{154906_16Nov26}

We compare the two contact-number structure factors $S_z(\bq)$ and
$S_z'(\bq)$ introduced in Sec.~\ref{140415_9Nov26}. Throughout this
appendix, we take $d\to\infty$ at fixed nonzero scaled wave number
$k=\sigma|\bq|/\sqrt{d}$.

The density stiffness derived in Sec.~\ref{140459_9Nov26} is of
order $d\hat{\varphi}$. Near GCP, where $\hat{\varphi}\sim\log d$,
its inverse gives
\begin{align*}
S_\rho(\bq)
\equiv \frac{1}{N}\ave{\rho_{\bq}\rho_{-\bq}}
= O\!\left(\frac{1}{d\log d}\right).
\end{align*}
The same order is obtained by evaluating the high-dimensional
hard-sphere liquid result,
$S_{\rm liq}(k)\simeq[1+d\hat{\varphi}e^{-k^2/2}]^{-1}$,
at the GCP density scale~\cite{schmid2010}.

For $\bq\ne0$, translational invariance gives $\ave{z_{\bq}}=0$.
The first term in Eq.~\eqref{151727_16Nov26} is therefore
\begin{align*}
S_z(\bq)=\frac{1}{N}\ave{z_{\bq}z_{-\bq}}=O(d),
\end{align*}
where the scaling follows from the calculation in the main text.
Since $\bar z=O(d)$, the density term is
\begin{align*}
\bar z^2 S_\rho(\bq)=O\!\left(\frac{d}{\log d}\right).
\end{align*}
The Cauchy--Schwarz inequality bounds the mixed term as
\begin{align*}
\frac{\bar z}{N}\left|\ave{\rho_{\bq}z_{-\bq}}\right|
&\leq \bar z\sqrt{S_\rho(\bq)S_z(\bq)}
=O\!\left(\frac{d}{\sqrt{\log d}}\right).
\end{align*}
Thus, at fixed $k$, the density-dependent terms are subleading,
and $S_z(\bq)$ and $S_z'(\bq)$ agree to leading order as $d\to\infty$.

\bibliography{reference}

\end{document}